\documentclass[epj]{svjour}
\usepackage[numbers,square,sort&compress]{natbib}
\usepackage{graphicx}
\usepackage{bm}                      % bold math
\usepackage{xcolor}                  % color
\usepackage{amssymb,amsmath}
\usepackage{multirow}
\def\bea{\begin{eqnarray}} \def\eea{\end{eqnarray}}
\def\beq{\begin{equation}} \def\eeq{\end{equation}}
\def\bal#1\eal{\begin{align}#1\end{align}}
\def\bse#1\ese{\begin{subequations}#1\end{subequations}}
\def\non{\nonumber}

\def\text{\mathrm}
\def\be{\beta}
\def\eps{\varepsilon}
\def\la{\Lambda}
\def\tl{\tilde\Lambda}
\def\ms{M_\odot}
\def\mmax{M_\text{max}}
\def\esnm{E_\text{SNM}}
\def\epnm{E_\text{PNM}}
\def\esym{E_\text{sym}}
\def\psym{p_\text{sym}}
\def\psnm{p_\text{SNM}}
\def\fm3{\;\text{fm}^{-3}}

\begin{document}

\title{Are nuclear matter properties correlated to neutron star observables ?}

\author{
Jin-Biao Wei\inst{1} \and
Jia-Jing Lu\inst{1,2} \and
G. F. Burgio\inst{1} \and
Zeng-Hua Li\inst{2} \and
H.-J. Schulze\inst{1}
}


\institute{
INFN Sezione di Catania, Dipartimento di Fisica e Astronomia,
Universit\`a di Catania,\\ Via S.~Sofia 64, I-95123 Catania, Italy
\and
Institute of Modern Physics, Key Laboratory of Nuclear Physics and
Ion-beam Application (MOE), Fudan University,\\ Shanghai 200433, P.R.~China}

\date{Received: date / Revised version: date}

\abstract{
We investigate properties of nuclear matter
and examine possible correlations with neutron star observables
for a set of microscopic nuclear equations of state
derived within the Brueckner-Hartree-Fock formalism
employing compatible three-body forces.
We find good candidates for a realistic nuclear EOS up to high density
and confirm strong correlations between neutron star radius,
tidal deformability, and the pressure of betastable matter.
No correlations are found with the saturation properties of nuclear matter.
}


\maketitle

%===============================================================================
\section{Introduction}
\label{s:intro}

Neutron star (NS) observations allow us to explore the equation of state (EOS)
of nuclear matter \cite{lattimer2016,2018Chap6}
at densities well beyond the ones available in terrestrial laboratories.
The nature of matter under conditions of extreme density and stability,
found only in the NS core, still remains an open question.
In particular, the mass and radius of NSs encode unique information
on the EOS at supranuclear densities.
Currently the masses of several NSs are known with good precision
\cite{mass,demorest2010,fonseca2016,heavy2,cromartie},
but the information on their radii is not very accurate
\cite{ozel16,gui2013,lat2014},
being more elusive than NS masses.
Particularly, a measurement of the radius with an error of about 1 km
could discriminate between soft and stiff EOSs,
as discussed in the current literature \cite{alfbur}.
For this purpose,
present observations of NICER \cite{nicer,nicer1,nicer2}
could in principle achieve an accuracy of about 2$\%$ for the radius,
whereas future planned missions like eXTP \cite{2019Watts}
will allow us to statistically infer their mass and radius
to within a few percent.
This information can be used to determine the EOS of the
matter in the NS interior,
and the nature of the forces between
fundamental particles under such extreme conditions.

The recent detection by the Advanced LIGO and Virgo collaborations
of gravitational waves emitted during the GW170817 NS merger event
\cite{merger,mergerl,mergerx}
has stimulated an intense research activity towards the understanding of the
nuclear matter EOS.
In particular, it provided important new insights on the structural properties
of these objects,
most prominently their masses and radii,
by means of the measurement of the tidal deformability \cite{hartle,flan},
and allowed to deduce upper \cite{merger,mergerl}
and lower \cite{radice} limits on it.

In this paper we compare the constraints on the nuclear EOS
obtained from heavy-ion collisions (HICs)
with those extracted from the analysis of the NS merger event GW170817.
We also examine possible correlations among properties
of nuclear matter close to saturation
with the observational quantities deduced from GW170817.
For this purpose we present recent calculations of NS structure
and their tidal deformability using various microscopic EOSs
for nuclear configurations.

The paper is organized as follows.
In Sec.~\ref{s:eos} we give a brief overview of the hadronic EOSs we are using,
and in Sec.~\ref{s:sat} we discuss their saturation properties
at normal nuclear density.
Constraints on the NS maximum mass are evaluated and
the tidal deformability as important NS observable is introduced in
Sec.~\ref{s:tid}.
In Sec.~\ref{s:cor} we investigate the compatibility of the EOSs
with constraints obtained from HIC data and the GW merger event,
and examine possible correlations between both.
In Sec.~\ref{s:end} we draw our conclusions.

%===============================================================================
\section{Equations of state}
\label{s:eos}

The theoretical description of nuclear matter under extreme density conditions
is a very challenging task.
Assuming that the most relevant degrees of freedom are nucleons,
thus neglecting other particles such as hyperons, kaons, or quarks,
the theoretical models can be either microscopic (ab-initio)
or phenomenological.

In this paper we use several EOSs based on the microscopic
Brueckner-Hartree-Fock (BHF) many-body theory \cite{jeu1976,baldo1999},
which provides a density expansion \cite{thl,thln,corr}
of the nuclear-matter binding energy
based on the use of realistic two-body forces.
It is well known that nucleonic three-body forces (TBF) are needed
in order to reproduce correctly the saturation properties of nuclear matter.
Currently a complete ab-initio theory of TBF is not available yet,
and therefore we adopt either phenomenological or microscopic models
\cite{glmm,uix3,zuo1,tbfnij}.
The microscopic BHF EOSs employed in this paper,
described in detail in Refs.~\cite{tbfnij,li08},
are based on different nucleon-nucleon potentials, namely
the Bonn B (BOB) \cite{bonn1,bonn2},
the Nijmegen 93 (N93) \cite{nij1,nij2},
and the Argonne $V_{18}$ (V18) \cite{v18}.
In the latter case, we also provide an EOS
obtained with the phenomenological Urbana model for describing TBF (UIX).
Useful parametrizations of these EOSs are given in the Appendix.

\begin{table*}[t]%..............................................................
\renewcommand{\arraystretch}{1.15}
\begin{center}
\caption{
\label{t:sat}
Saturation properties and NS observables predicted by the considered EOSs.
Experimental nuclear parameters and observational data
from different sources
are also listed for comparison.
See text for details.}
%\medskip
%\begin{ruledtabular}
\begin{tabular}{ccccccccc}
\hline\hline
 EOS    & $\rho_0[\fm3]$ & $-E_0$[MeV] & $K_0$[MeV] & $S_0$[MeV] & $L$[MeV]
        & $\mmax[\ms]$ & $\la_{1.4}$ & $R_{1.4}$[km] \\
\hline
 BOB     & 0.170  & 15.4 & 238 & 33.7 & 70 & 2.50 & 570 & 12.9 \\
 V18     & 0.178  & 13.9 & 207 & 32.3 & 67 & 2.36 & 442 & 12.3 \\
 N93     & 0.185  & 16.1 & 229 & 36.5 & 77 & 2.25 & 473 & 12.7 \\
 UIX     & 0.171  & 14.9 & 171 & 33.5 & 61 & 1.96 & 309 & 11.8 \\
 FSS2CC  & 0.157  & 16.3 & 219 & 31.8 & 52 & 1.94 & 295 & 11.8 \\
 FSS2GC  & 0.170  & 15.6 & 185 & 31.0 & 51 & 2.08 & 262 & 11.5 \\
 DBHF    & 0.181  & 16.2 & 218 & 34.4 & 69 & 2.31 & 681 & 13.1 \\
 APR     & 0.159  & 15.9 & 233 & 33.4 & 51 & 2.20 & 274 & 11.6 \\
 LS220   & 0.155  & 15.8 & 219 & 27.8 & 68 & 2.04 & 542 & 12.9 \\
 SFHO    & 0.157  & 16.2 & 244 & 32.8 & 53 & 2.06 & 334 & 11.9 \\
\hline
 Exp.   & $\sim$ 0.14--0.17 & $\sim$ 15--16 & 220--260 & 28.5--34.9 & 30--87
        & $>2.14^{+0.10}_{-0.09}$
        & 70--580 & 10.5--13.3 \\
%        & $190^{+390}_{-120}$ & $11.9^{+1.4}_{-1.4}$ \\
 Ref.   & \cite{margue2018a} %audi2012,2013ADNDT..99...69A}
        & \cite{margue2018a} %brown,browne}
        & \cite{shlomo,pieka}
        & \cite{lihan,oertel} & \cite{lihan,oertel} & \cite{cromartie}
        & \cite{mergerl} & \cite{mergerl} \\
\hline\hline
\end{tabular}
%\end{ruledtabular}
\end{center}
%\vspace*{2cm}
\end{table*}%...................................................................

In the same theoretical framework,
we also studied an EOS based on a potential model which includes
explicitly the quark-gluon degrees of freedom, named fss2
\cite{2014PhRvL.113x2501B,2015PhRvC..92f5802F}.
This reproduces correctly the saturation point of symmetric matter
and the binding energy of few-nucleon systems
without the need of introducing TBF.
In the following those EOSs are labeled FSS2CC and FSS2GC,
indicating two different prescriptions of solving the BHF equations.

We compare these BHF EOSs with the often-used results of the
Dirac-BHF method (DBHF) \cite{dbhf3},
which employs the Bonn~A potential,
and the APR EOS \cite{apr1998}
based on the variational method and the $V_{18}$ potential.
For completeness,
we also use the well-known phenomenological LS220 \cite{ls}
and SFHo \cite{sfh} EOSs,
both based on the relativistic mean-field (RMF) approach
for the high-density part.

The BHF method provides EOSs for homogeneous nuclear matter,
$\rho > \rho_t \approx 0.08\,\text{fm}^{-3}$.
For the low-density inhomogeneous crustal part
we adopt the well-known Negele-Vautherin EOS
\cite{NV} for the inner crust in the medium-density regime
($0.001\,\text{fm}^{-3} < \rho \lesssim 0.08\,\text{fm}^{-3}$),
and the ones by Baym-Pethick-Sutherland \cite{bps}
and Feynman-Metropolis-Teller \cite{fey} for the outer crust
($\rho < 0.001\,\text{fm}^{-3}$).
%We mention that for the calculation of stellar structure we used the EOSs
%of Refs.~\cite{fey,bps} for the outer crust
%and \cite{NV} for the inner crust.
The transition density $\rho_t$ is adjusted to provide a smooth transition
of pressure and energy density between both branches of the betastable EOS.
The NS-mass domain that we are interested in,
is hardly affected by the structure
of this low-density transition region
and the crustal EOS \cite{bhfls}:
The choice of the crust model can influence the predictions of
radius and related deformability to a small extent,
of the order of $1\%$ for $R_{1.4}$ \cite{bhfls,2014Bur_Cent,fortin},
which is negligible for our purpose.
Even neglecting the crust completely,
NS radius and deformability do not change dramatically \cite{tsang19}.

%===============================================================================
\section{EOS properties at saturation}
\label{s:sat}

It is very important that any property of the adopted EOS
can be tested at the saturation density $\rho_0\approx0.17\fm3$
of symmetric nuclear matter (SNM)
[$N=Z$, being $N (Z)$ the neutron (proton) number],
where information from laboratory data on finite nuclei is available.
In general, in the vicinity of the saturation point
the binding energy per nucleon can be expressed in terms of the
density parameter
$x \equiv (\rho-\rho_0)/3\rho_0$
and the asymmetry parameter
$\delta \equiv (N-Z)/(N+Z)$
as
\bea
 E(\rho,\delta) &=& \esnm(\rho) + \esym(\rho) \delta^2 \:,
\label{e:ea}
\\
 \esnm(\rho) &=& E_0 + \frac{K_0}{2} x^2 \:,
\\
 \esym(\rho) &=& S_0 + L x + \frac{K_\text{sym}}{2} x^2 \:,
\label{e:ebulk}
\eea
where
$K_0$ is the incompressibility and
$S_0 \equiv E_\text{sym}(\rho_0)$
is the symmetry energy coefficient at saturation,
and the parameters $L$ and $K_\text{sym}$
characterize the density dependence of the symmetry energy around saturation.
The incompressibility $K_0$ gives the curvature of $E(\rho)$ at $\rho=\rho_0$,
whereas $S_0$ determines the increase of the energy per nucleon
due to a small asymmetry $\delta$.
These parameters are defined as
\bea
 K_0 &\equiv& 9\rho_0^2  \frac{d^2\esnm}{d\rho^2}(\rho_0) \:,
\\
 S_0 &\equiv& \frac{1}{2} \frac{\partial^2E}{\partial\delta^2}(\rho_0,0)
 \approx E_\text{PNM}(\rho_0) - \esnm(\rho_0) \:,
\\
 L &\equiv& 3 \rho_0 \frac{d\esym}{d\rho}(\rho_0) \:,
\\
 K_\text{sym} &\equiv& 9 \rho_0^2 \frac{d^2\esym}{d\rho^2}(\rho_0) \:.
\eea

Properties of the various considered EOSs are listed in Table~\ref{t:sat},
namely, the value of the saturation density $\rho_0$,
the binding energy per particle $E_0$,
the incompressibility $K_0$,
the symmetry energy $S_0$
(note that we use the second definition
involving the energy of pure neutron matter (PNM) for the values in the table),
and its derivative $L$ at $\rho_0$.
The curvature of the symmetry energy $K_\text{sym}$ is only loosely known to
be in the range of
$-400\,\text{MeV} \lesssim K_\text{sym} \lesssim 100\,\text{MeV}$
\cite{tews2017,2017arXiv170402687Z},
and therefore will not be examined in this paper.
In Table I we have also included the experimental ranges for the nuclear
parameters and the data available so far from astrophysical observations.
We notice that all the adopted EOSs agree fairly well with the empirical values.
Marginal cases are the
slightly too low $E_0$ and $K_0$ for V18,
too small/large $S_0$ for LS220/N93,
and too low $K_0$ for UIX and FSS2GC.
The $L$ parameter does not exclude any of the EOSs.

%===============================================================================
\section{Neutron star structure and tidal deformability}
\label{s:tid}

A very important constraint
to be fulfilled by the different EOSs
(assuming a purely nucleonic composition of NS matter)
is the value of the maximum NS mass,
which has to be compatible with the observational data
\cite{demorest2010,fonseca2016,heavy2},
in particular the recent lower limit
$\mmax>2.14^{+0.10}_{-0.09}$ \cite{cromartie}.
In general relativity, the maximum mass is calculated by solving the
Tolman-Oppenheimer-Volkoff (TOV) equations
for pressure $p$ and enclosed mass $m$ of a static NS configuration
\bea
  {dp\over dr} &=& -{ m \eps \over r^2 }
%\nonumber\\ && \times
  {  \left( 1 + {p/\eps} \right) \left( 1 + {4\pi r^3 p/m} \right)
  \over 1-2m/r } \:,
\label{e:tov1}
\\
  {dm \over dr} &=& 4 \pi r^2 \eps \:.
\label{e:tov2}
\eea
The only input required is the EOS $\eps(p)$.
As shown in Table~\ref{t:sat},
many EOS models give compatible values of the maximum mass,
apart from UIX, FSS2CC, and marginally LS220.
We notice that recent analyses of the GW170817 event
indicate also an upper limit of the maximum mass of about
2.2--2.3$\,\ms$
\cite{shiba17,marga17,rezz18,shiba19},
with which several of the microscopic EOSs (V18, N93, DBHF, and APR)
would also be compatible.

%We mention that the BHF theory was also extended in order to include hyperons,
%which might appear in the core of a NS,
%but the corresponding EOSs turn out to be very soft,
%with too low NS maximum masses,
%$M<1.7\,\ms$ ($\ms\approx2\times10^{33}$g) \cite{mmy1,mmyy,mmy2},
%well below the current observational limit.

The tidal deformability $\lambda$,
or equivalently the tidal Love number $k_2$ of a NS
\cite{hinder2008,hinder2009,hinder2010},
has recently been acknowledged to provide valuable information and constraints
on the related EOS.
More specifically, the Love number
%\begin{widetext}
\bea
 k_2 &=& \frac{3}{2}\frac{\lambda}{R^5} = \frac{3}{2} \be^5 \la
 = \frac{8}{5}\frac{\be^5 z}{F} \:,
\label{e:l}
\\
 && z \equiv  (1-2\be)^2 [2-y_R+2\be(y_R-1)] \:,
\non\\
 && F \equiv  6\be(2-y_R) + 6\be^2(5y_R-8) + 4\be^3(13-11y_R)
\non\\\non
   &&\hskip6mm  +\, 4\be^4(3y_R-2) + 8\be^5(1+y_R) + 3z\ln(1-2\be) \:
\eea
%\end{widetext}
with $\la\equiv\lambda/M^5$ and $\be\equiv M/R$ being the compactness,
can be obtained by solving the TOV equations~(\ref{e:tov1},\ref{e:tov2})
along with the following first-order differential equation
\cite{eos},
\bea
  {dy \over dr} &=& -\frac{y^2}{r} - \frac{y-6}{r-2m} - rQ\:,
\non\\
  && Q \equiv 4\pi \frac{(5-y)\eps+(9+y)p+(\eps+p)/c_s^2}{1-2m/r}
\non\\
  && \phantom{Q \equiv} - \Bigg[ \frac{2(m+4\pi r^3 p)}{r(r-2m)} \Bigg]^2,
\label{e:tov3}
\eea
with the EOS $\eps(p)$ as input,
$c_s^2=d\!p/d\eps$ the speed of sound,
and boundary conditions given by
\beq
 [p,m,y](r=0) = [p_c,0,2] \:,
\eeq
being $y_R\equiv y(R)$,
and the mass-radius relation $M(R)$
provided by the condition $p(R)=0$ for varying $p_c$.

For an asymmetric binary NS system,
$(M,R)_1+(M,R)_2$, with mass asymmetry $q=M_2/M_1$,
and known chirp mass
\beq
 M_c = \frac{(M_1 M_2)^{3/5}}{(M_1+M_2)^{1/5}} \:,
\eeq
the average tidal deformability is defined by
\beq
 \tl = \frac{16}{13}
 \frac{(1+12q)\la_1 + (q+12)\la_2}{(1+q)^5}
\label{e:lq}
\eeq
with
\beq
 \frac{[M_1,M_2]}{M_c} =
 \frac{297}{250} (1+q)^{1/5} [q^{-3/5},q^{2/5}] \:.
\eeq

From the analysis of the GW170817 event \cite{mergerx},
a value of $M_c/\ms=1.186{+0.001\atop-0.001}$ was obtained,
corresponding to $M_1=M_2=1.36\,\ms$ for a symmetric binary system,
$q=0.73-1$ and $\tl<730$ from the phase-shift analysis of the observed signal.
A lower limit, $\tl>400$,
was deduced from a multimessenger analysis of the GW170817 event
combined with an analysis of the UV/optical/infrared counterpart
with kilonova models \cite{radice}.
Recently, this value was updated to
$\tl\gtrsim300$ \cite{radice2,coughlin},
although this lower bound has been disputed \cite{kiuchi}.

We report in Table~\ref{t:sat} the value of the tidal deformability
for a $1.4\,\ms$ NS, and the corresponding radius $R_{1.4}$.
It turns out that,
requiring both NSs to have the same EOS,
leads to constraints $70 < \la_{1.4} < 580$ and
$10.5 < R_{1.4} < 13.3$ km \cite{mergerl}.
We notice that the values calculated for all the adopted EOSs,
and displayed in Table~\ref{t:sat},
lie in the above intervals with the exception of DBHF.

%===============================================================================
%\section{Results and discussion}
\section{Constraints and Correlations}
\label{s:cor}

%-------------------------------------------------------------------------------
\subsection{Constraints on the symmetry energy}
%around saturation density and above}

\begin{figure}[t]%..............................................................
\vspace{-3mm}\hspace{2mm}
\centerline{\includegraphics[scale=0.38]{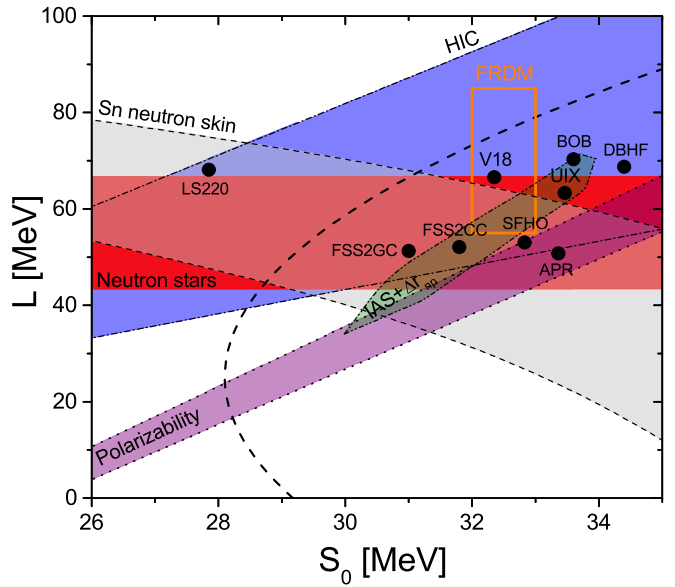}}
\vspace{-8mm}
\caption{
Correlations between symmetry energy $S_0$ and its slope $L$
at the saturation density.
The markers represent the predictions of the considered EOSs.
See text for details on the various constraints.}
\label{f:fig1}
\end{figure}

An important test for the EOS has to do with the symmetry energy,
for which the experimental constraints are abundant at saturation density
(see, e.g., \cite{tsang2012,latlim2013,latste2014b}).
We show in Fig.~\ref{f:fig1} a set of different experimental constraints
together with the values of $(S_0,L)$ predicted by the various
theoretical models considered in this paper.
More in detail,\\
-- the label ``HIC" (blue region) corresponds to the constraints
inferred from the study of isospin diffusion in HICs
\cite{2009PhRvL.102l2701T};\\
-- the label ``Polarizability" (violet region) represents the constraints
on the electric dipole polarizability deduced in \cite{rocavin2015};\\
-- the label ``Sn neutron skin" (grey region) indicates the constraints deduced
from the analysis of neutron skin thickness in Sn isotopes \cite{chen2010};\\
-- the label ``FRDM" (rectangle) corresponds to the values of $S_0$ and $L$
inferred from finite-range droplet mass model calculations \cite{moller2012};\\
-- the label ``IAS $+ \Delta r_{\rm np}$'' (green diagonal region)
indicates the isobaric-analog-state (IAS) phenomenology
combined with the skin-width data, and
represents simultaneous constraints by Skyrme-Hartree-Fock calculations
of the IAS and the $^{208}$Pb neutron-skin thickness \cite{danlee2014};\\
-- the horizontal band (in red color) labeled ``Neutron stars'' is obtained
from a Bayesian analysis of mass and radius measurements of NSs
by considering the 68\% confidence values for $L$ ~\cite{stelat2013};\\
-- the dashed curve is the unitary gas bound on symmetry energy parameters
derived in Ref.~\cite{tews2017}:
only values of $(S_0,L)$ to the right of the curve are permitted.

%...................................................................

All considered constraints are not simultaneously fulfilled
in any area of the parameter space,
and this is probably due to the model dependencies
that influence the derivation of constraints from the raw data,
besides the current uncertainties in the experimental measurements.
Given this situation,
at the moment no definitive conclusion can be drawn and,
except for models predicting values of the symmetry energy parameters
outside the limits given in Table~\ref{t:sat}
(like the LS220 or the N93 EOS),
no theoretical models can be ruled out a priori on this basis.

\begin{figure}[t]%..............................................................
\vspace{-1mm}\hspace{2mm}
\centerline{\includegraphics[scale=0.37]{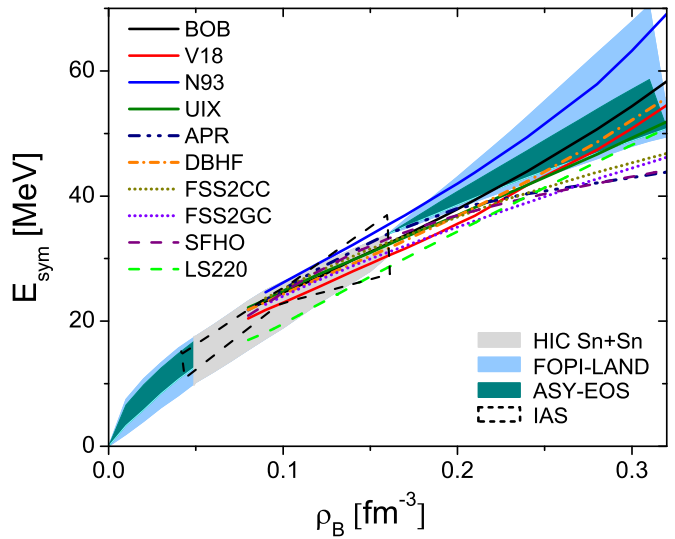}}
\vspace{-6mm}
\caption{
Symmetry energy vs.~baryon density for all considered EOSs.
The blue, green, and grey bands, as well as the dashed box,
represent experimental data as described in the text.}
\label{f:fig2}
\end{figure}%...................................................................

A further crucial point in the understanding of the nuclear symmetry energy
is its high-density behavior,
which is among the most uncertain properties of dense neutron-rich matter.
Its accurate determination has significant consequences in understanding
not only the reaction dynamics of heavy-ion reactions,
but also many interesting phenomena in astrophysics,
such as the explosion mechanism of supernovae and the properties of NSs.
In fact several aspects of the NS structure and dynamics depend crucially on the
symmetry energy, e.g.,
the composition and the onset of the direct Urca cooling reaction,
which is a threshold process dependent on the proton fraction
controlled by the symmetry energy.

A big experimental effort has been devoted during the last few years
to constrain the high-density symmetry energy
using various probes in HICs at relativistic energies.
Fig.~\ref{f:fig2} displays some constraints deduced for the density dependence
of the symmetry energy
from the ASY-EOS data \cite{2016PhRvC..94c4608R} (green band)
and the FOPI-LAND result \cite{2011PhLB..697..471R} (blue band)
as a function of the density.
The results of Ref.~\cite{2009PhRvL.102l2701T}
are reported in the grey area (HIC Sn+Sn),
whereas the dashed contour labeled by IAS shows the results
of Ref.~\cite{danlee2014}.
We observe that the experimental results exhibit a monotonically increasing
behavior with increasing density,
and that several microscopic EOSs turn out to be compatible with experiments,
except LS220 around saturation density,
whereas N93, FSS2CC, and FSS2GC above the saturation density
are only marginally compatible with the data.

\begin{figure*}[t]%.............................................................
\vspace{-4mm}
\centerline{\includegraphics[scale=0.65]{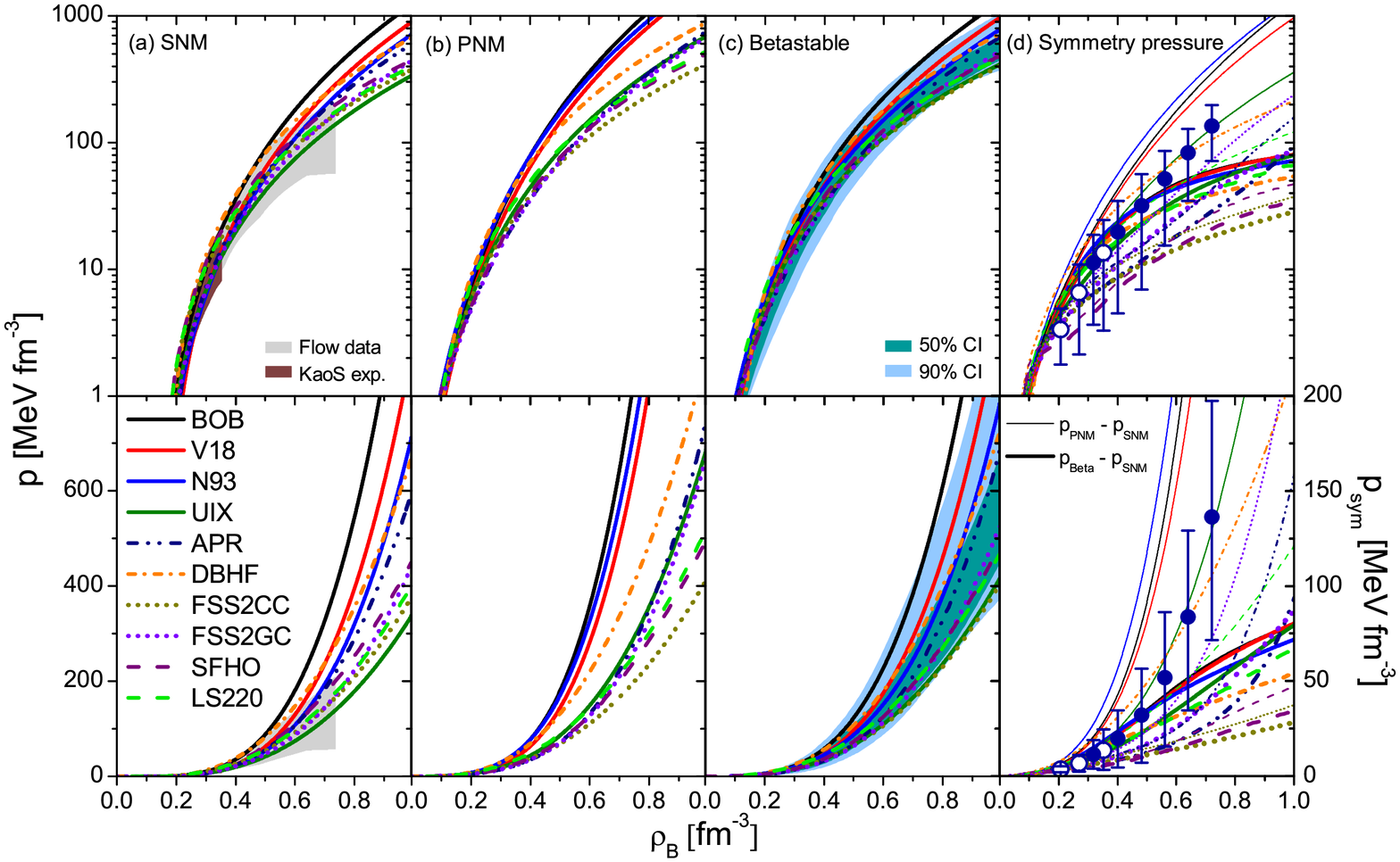}}
\vspace{-3mm}
\caption{
Pressure vs.~baryon density
for the considered EOSs
on a logarithmic (upper row) or linear (lower row) scale for
(a) symmetric matter,
(b) pure neutron matter,
(c) beta-stable matter, and
(d) the symmetry pressure.
In (a) constraints derived from HIC data are reported
as brown band (KaoS experiment) and grey band (Flow data).
In (c) the GW170817 constraints \cite{mergerl} are reported.
The markers in (d) are from the data analysis of
Ref.~\cite{2019arXiv190107673T}.
The thick curves represent the compatible quantity $p_\beta-\psnm$,
whereas the actual symmetry pressure  $p_\text{PNM} - \psnm$
is shown by thin curves.
See text for more details.}
\label{f:fig3}
\end{figure*}%..................................................................

\begin{figure}[t]%..............................................................
\vspace{-17mm}
\centerline{\hspace{58mm}\includegraphics[scale=0.65]{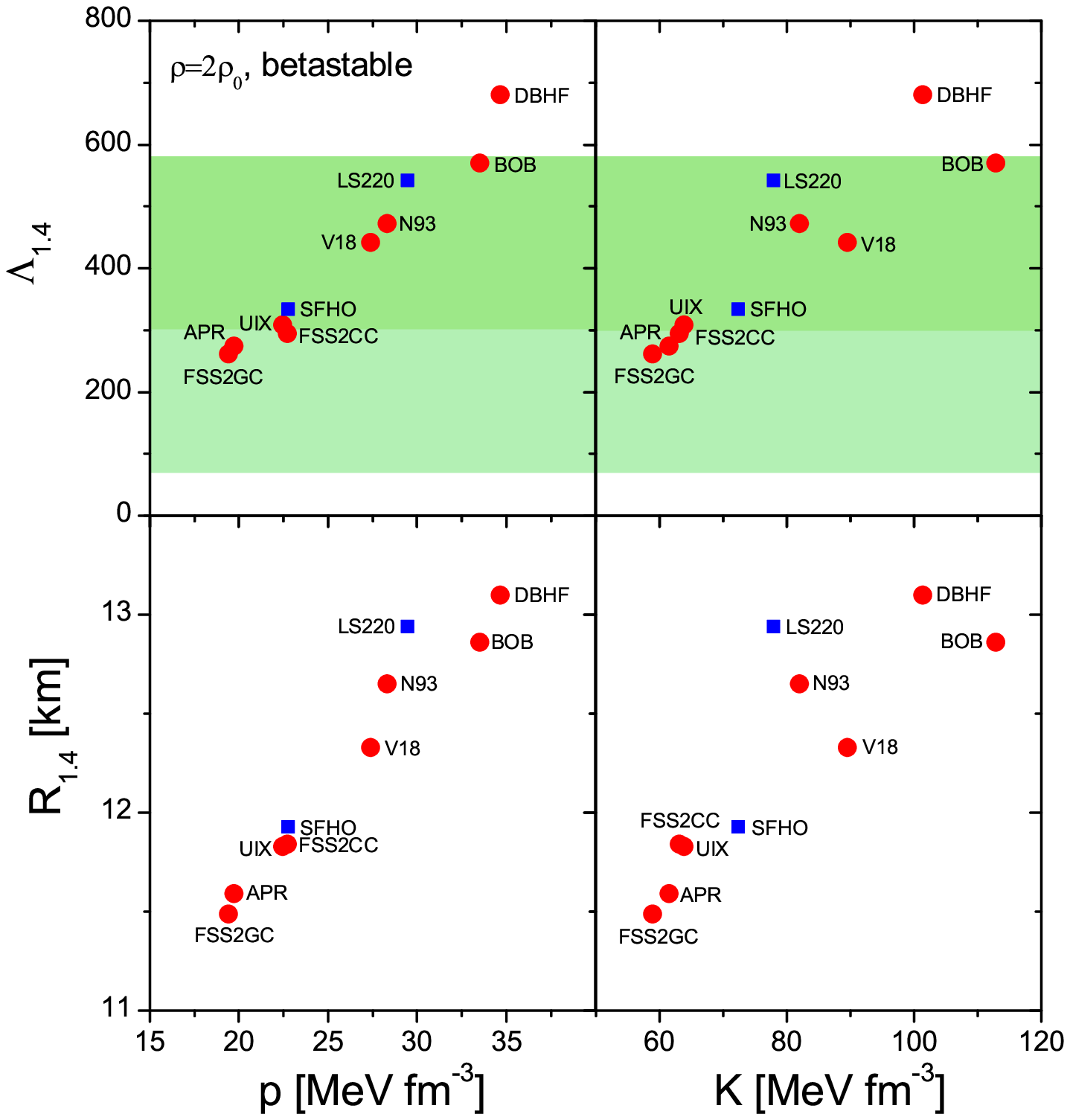}}
\vspace{-9mm}
\caption{
Tidal deformability (upper panels)
and radius (lower panels)
of a $1.4\,\ms$ NS
vs.~the pressure (left panels)
and the incompressibility $K$ (right panels)
of betastable matter
at twice the saturation density.
The light and dark shaded bands
in the upper row represent the limits derived in
\cite{mergerl,radice2}, respectively.}
\label{f:fig4}
\end{figure}%...................................................................

%-------------------------------------------------------------------------------
\subsection{Constraints on the EOS from heavy-ion collisions
and gravitational waves}

The extraction of the gross properties of the nuclear EOS
from HIC data has been one of the main objectives
in terrestrial nuclear experiments in the last two decades.
In fact HICs at energies ranging from few tens
to several hundreds MeV per nucleon produce heavily compressed nuclear matter
with subsequent emission of nucleons and fragments of different sizes.
The experimental analysis has been performed using the transverse flow
as an observable,
since it strongly depends on the pressure developed in the interaction zone
of the colliding nuclei at the moment of maximum compression.
The fireball density reached during the collision can also be probed
by subthreshold $K^+$ production,
since this depends on its incompressibility,
as shown by the data collected by the KaoS collaboration \cite{kaos}.
A combined flow and kaon production analysis
was presented in Ref.~\cite{daniel2002},
where a region in the pressure vs.~density plane was identified,
through which a compatible EOS should pass.

That analysis is displayed in Fig.~\ref{f:fig3} (left panels)
as a grey box for the flow data by the FOPI collaboration \cite{fopi},
and as a brown box for the KaoS collaboration \cite{kaos}.
Those results point in the direction of a soft EOS,
with values of the incompressibility $K$ in the range
$180 \leq K \leq 250$ MeV close to the saturation density.
We observe that almost all considered EOSs
are compatible with the experimental data,
except the BOB, V18, and DBHF EOS,
which are too stiff at large density,
where the analysis could however
be less reliable due to the possible appearance
of other degrees of freedom besides nucleons.
Such densities are actually never reached in HICs.
For completeness, we display in the central panels (b)
the pressure for the PNM case.

The EOS governs also the dynamics of NS mergers.
In fact, the possible scenarios of a prompt or delayed collapse
to a black hole or a single NS, following the merger, do depend on the EOS,
as well as the amount of ejected matter which undergoes nucleosynthesis
of heavy elements.
During the inspiral phase,
the EOS strongly affects the tidal polarizability $\la$, Eq.~(\ref{e:l}).
The first GW170817 analysis for a $1.4\,\ms$ NS \cite{merger}
gave an upper limit of $\la_{1.4}<800$,
which was later improved to $\la_{1.4}=190^{+390}_{-120}$
by assuming that both NSs feature the same EOS \cite{mergerl}.
In this new analysis,
the values of the pressure as a function of density were extracted,
and those are displayed as colored areas in Fig.~\ref{f:fig3}(c),
in which the blue (green) shaded region corresponds to the 90$\%$ (50$\%$)
posterior confidence level.
We notice that almost all EOSs turn out to be compatible with the
GW170817 data at density $\rho>2\rho_0$,
with BOB in marginal agreement at large density.
This constraint
combined with the recent observation of the new maximum mass
$M=2.14^{+0.10}_{-0.09}\ms$ of PSR J0740+6620 \cite{cromartie}
represents at the moment the strongest test for any EOS model.
In our case, the V18 EOS appears the most compatible with both data sets.
This point has also been discussed in the framework of phenomenological EOSs
\cite{2019Zhang},
where the combined data help to constrain the range of values of the stiffness
of isospin-symmetric nuclear matter.
A further comparison of HIC data with GW observations
can be found in Ref.~\cite{2018arXiv180706571T}.

Another interesting quantity to consider is the so-called symmetry pressure,
\beq
 \psym(\rho) = \rho^2 \frac{dE_\text{sym}(\rho)}{d\rho}
 \approx p_\text{PNM}(\rho) - \psnm(\rho) \:,
\eeq
[the last equation is valid in case of the quadratic approximation
Eq.~(\ref{e:ea})],
which adds to the pressure of an isospin-symmetric system with $N=Z$.
Its contribution is very important because it is related
to the poorly known symmetry energy at large density,
and plays a big role in the determination of the proton fraction, for instance,
crucial for NS cooling simulations.
More precisely, the pressure of betastable matter (including electrons)
with asymmetry $\delta(\rho)=(\rho_n-\rho_e)/\rho$
is given by \cite{lat2000,zhang19,2019Zhang}
\beq
 p_\beta(\rho)
 =  \psnm(\rho) +  \delta^2 \psym(\rho)
 + \frac{\delta(1-\delta)}{2} \rho\esym(\rho) \:.
\eeq
For small electron fractions one has $\delta\approx 1$ and
\beq
 p_\text{sym} \approx p_\beta - p_\text{SNM} \:.
\label{e:psym}
\eeq
This quantity is displayed in Fig.~\ref{f:fig3}(d)
using thick curves.
The markers with error bars are the corresponding results of the
analysis performed in Ref.~\cite{2019arXiv190107673T},
where a subtraction procedure has been proposed between
the kaon data (white dots) and flow data (blue dots) for SNM,
both displayed in panel (a),
and the GW170817 event constraints shown in panel (c),
assuming matter in beta-stable condition.
We see that the symmetry pressure increases rapidly with the baryon density,
as many microscopic EOSs predict,
except at densities above $\rho\gtrsim 0.7\fm3$,
where most EOSs show a saturating behavior
and thus a different trend with respect to the analysis of the (Flow) data.
However, such high densities are never actually reached in HICs.

We also point out that the true symmetry pressure
$p_\text{sym} \approx p_\text{PNM} - p_\text{SNM}$
[thin curves in Fig.~\ref{f:fig3}(d)]
can be substantially larger than the approximation Eq.~(\ref{e:psym}),
if one takes properly into account the EOS for PNM,
which is significantly stiffer than the betastable EOS
for most considered models,
compare Figs.~\ref{f:fig3}(b) and (c).

%-------------------------------------------------------------------------------
\subsection{Correlations between neutron star and nuclear matter observables}

\begin{figure*}[t]%.............................................................
\vspace{-5mm}
\centerline{\includegraphics[scale=0.7]{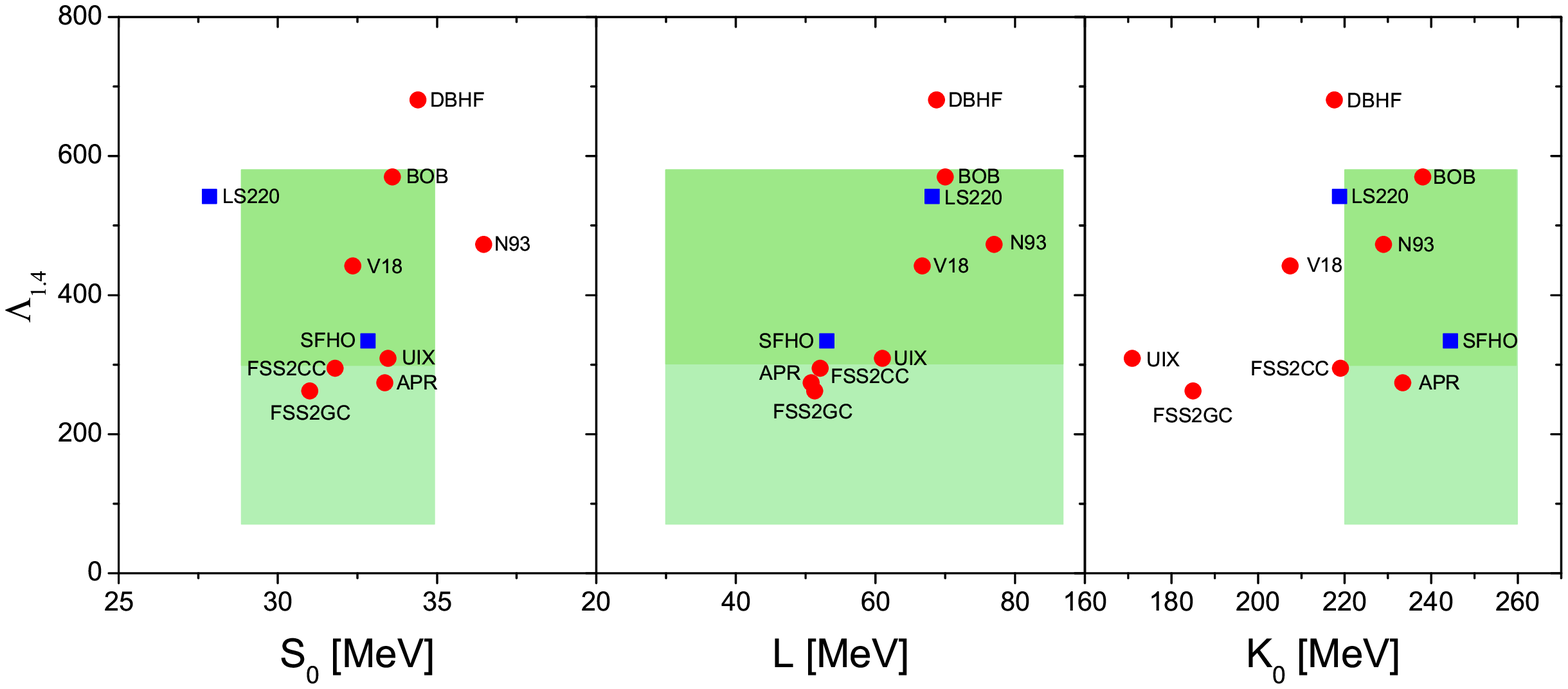}}
\vspace{-2mm}
\caption{
The tidal deformability of a $1.4\,\ms$ NS
as a function of the symmetry energy (left panel),
its derivative $L$ (middle panel),
and the incompressibility $K_0$ at saturation density
$\rho_0$ for all the considered EOSs.
The shaded areas represent the limits listed in Table~\ref{t:sat}.
}
\label{f:fig5}
\end{figure*}%..................................................................

In order to better understand the properties of nuclear matter,
it would be very interesting to find correlations between GW170817 observations
and microscopic constraints from nuclear measurements,
as the ones just discussed.
For this purpose,
the limits derived for the tidal deformability could be very valuable.

We remind that the LIGO-Virgo collaborations performed a Bayesian analysis
of the GW data,
assuming that each star may have different EOSs \cite{merger},
thus deducing a bound on $\la_{1.4} \leq 800$ for a $1.4\,\ms$ NS.
This bound rules out very stiff EOSs.
Based on this result, the authors of Ref.~\cite{annala17}
combined an EOS at low densities built with the chiral effective theory
with perturbative QCD results at large densities,
and constrained  $R_{1.4} < 13.6 \,$km \cite{fatto17}.
Similar results were obtained also in
Refs.~\cite{fatto17,most18,lim18,raithel18,malik18}.
The improved analysis of the GW170817 merger event,
performed using more realistic waveform models
and assuming the same EOS for both stars,
provided $70 < \la_{1.4} < 580$ at $90\%$ confidence level \cite{mergerl},
which allowed to tighten the upper limit to $R_{1.4} < 12.9$~km
in a recent analysis \cite{nandi19} in the framework of RMF models.
%We use this constraint in the discussion of our results.
The interpretation of the GW170817 event allowed to establish also
{\em lower} limits on the NS radius, e.g.,
the measurement of the neutron skin thickness of $\rm^{208}Pb$
by the PREX collaboration \cite{abra12}
gives $R_{1.4} > 12.55 \,$km \cite{fatto17},
and this was confirmed by similar recent analyses \cite{most18,lim18}.

The source of GW170817 also released a short gamma-ray burst, GRB170817A,
and a kilonova, AT2017gfo, generated by the mass ejected from the merger,
and this was found to provide constraints on the EOS as well.
In particular \cite{radice},
a limit $\tl > 400$
on the average tidal deformability, Eq.~(\ref{e:lq}),
was deduced in order to eject material heavier than $0.05\,\ms$,
as required by the high luminosity of AT2017gfo.
This constraint could indicate that $R_{1.4}\gtrsim 12\,$km,
which was used in Refs.~\cite{most18,lim18,malik18,drago4}
in order to constrain properties of nuclear matter.
This lower limit has been recently loosened to $\tl \gtrsim 300$ \cite{radice2},
although in Ref.~\cite{kiuchi} it has been considered of limited significance.

We remind the reader that there are no precise simultaneous measurements
of a NS mass and radius yet.
To date, several different astrophysical measurements of NS radii
have been attempted,
and they still suffer of big uncertainties \cite{ozel16},
especially related to the atmosphere composition.
After the first inference of radii from the thermal emission
from accreting NSs in quiescent low-mass X-ray binaries (QLMXBs)
\cite{2010Ozel,stelat2013},
radii in the range 10.1--11.1~km have been deduced
from new measurements with a larger sample of sources \cite{2016Ozel}.
They are smaller than the lower limits derived from GW170817 observations.
It has been suggested that this discrepancy (if confirmed)
might be solved in the two-families or twin-star scenarios,
in which small and big stars of the same mass
could coexist as hadronic and quark matter stars
\cite{drago2,drago3,drago4,pascha}.
On the other hand,
a different analysis was proposed in Ref.~\cite{latste2014b},
where the radius estimates were inferred from
photospheric radius expansion (PRE) bursts and thermal emissions from
QLMXBs and isolated NSs.
In that case, typical radii, i.e., for 1.2--1.8$\,\ms$ stars,
lie in the range 11.1--12.8~km,
fully compatible with GW170817.
These existing discrepancies could be overcome by the in-progress NASA
Neutron Star Interior ExploreR (NICER) mission \cite{nicer,nicer1,nicer2},
which will hopefully provide in the near future radii measurements
with uncertainties as small as 0.5 km.

We now turn to the discussion of our results.
In Ref.~\cite{lat2000} it was found that
for a wide choice of nuclear EOSs
the NS radius is strongly correlated with the pressure of betastable matter
$p=\rho^2dE_\beta/d\rho$
at a density $\rho\approx2\rho_0$.
According to the previous discussion
this correlation is then also valid between $\la_{1.4}$ and pressure.
This is also confirmed for our set of EOSs as shown
in Fig.~\ref{f:fig4} (left panels),
whereas weaker correlations appear with the incompressibility
$K=9\rho^2d^2E_\beta/d\rho^2$
under the same conditions, displayed in the right panels.

For completeness, we have calculated the correlation factors
\beq
 r(x,y) = \frac{1}{n-1}
 \frac{\sum_x \sum_y (x-\bar{x}) (y-\bar{y})}{s_x s_y} \:,
\eeq
where $n$ is the number of data pairs,
$\bar{x}$ and $\bar{y}$ are the sample means
of all the $x$ and $y$ values, respectively;
and $s_x$ and $s_y$ are their standard deviations.
In our case the results are
\bal
 r(p_{},\la_{1.4}) &= 0.982 \ ,
 &r(p_{},R_{1.4}) = 0.971 \:,
\\
 r(K,\la_{1.4}) &= 0.885 \ ,
 &r(K,R_{1.4}) = 0.846 \:,
\eal
which confirm the above statements,
i.e., stronger (weaker) correlations among
pressure $p$, $\la_{1.4}$, and $R_{1.4}$
($K$, $\la_{1.4}$, and $R_{1.4}$).

The green bands displayed in the upper panels represent the limits
on $\la_{1.4}$ derived in \cite{mergerl,radice2},
in particular the lower limits, i.e.,
$\la_{1.4}=190^{+390}_{-120}$ \cite{mergerl} (light green) and
$\la_{1.4}>300$ \cite{radice2,coughlin} (dark green),
are important for the determination of the radius,
which corresponds to
$R_{1.4}=11.9^{+1.4}_{-1.4}$ km in the former case, and
$R_{1.4}=12.2^{+1.0}_{-0.8}\pm 0.2$ km in the latter one.
%We notice that these values of $\la$ were recently updated to
%$\tilde\la=300^{+500}_{-190}$ \cite{mergerx}.
For completeness, we have checked whether this correlation
applies also to NS masses different from $1.4\,\ms$,
but it becomes slightly weaker with increasing NS masses.
Thus the determination of the tidal deformability or the NS radius
could put constraints on the pressure and the symmetry pressure
at twice the saturation density \cite{lat2000,eos}.
The current limits exclude only the DBHF EOS due to its
too high $\la_{1.4}$ value.

Following the same philosophy,
we have tried to find correlations between NS observables
and properties of SNM around saturation density.
Results are displayed in Fig.~\ref{f:fig5}
(the green bands display the same conditions as in Fig.~\ref{f:fig4}),
where the tidal deformability of a $1.4\,\ms$ NS is reported
as a function of the symmetry energy $S_0$ (left panel),
its slope $L$ (middle panel),
and the incompressibility $K_0$ (right panel),
all taken at saturation density.
Apparently no evident correlations between the tidal deformability
and $S_0$ and $K_0$ do exist,
whereas some degree of correlation is found with $L$,
as confirmed by the corresponding correlation factors
\beq
 r([S_0,K_0,L],\la_{1.4}) = [0.128,0.300,0.808] \:.
\eeq

Similar results were found also in
Refs.~\cite{PhysRevC.87.015806,malik18,perot19,tsang19},
with several EOSs based on the RMF model
and the Skyrme-Hartree-Fock approach.

%===============================================================================
\section{Summary}
\label{s:end}

We conclude that among the BHF models analyzed here,
the V18 and N93 could be good candidates for a
realistic description of the nuclear EOS up to very high density.
They fulfill nearly all current experimental and observational constraints
discussed in this article,
in particular the novel constraints on the tidal deformability imposed
by GW170817.
We would like to emphasize that these are not phenomenological EOSs,
but they have been constructed in a microscopic way
from nuclear two-body potentials and compatible three-body forces.
The last issue imposed in fact strong conditions on their construction,
due to which reason a perfect reproduction of all
current constraints is not achieved,
but was also not attempted.
We stress in particular that the predicted maximum mass values $\approx2.3\,\ms$
could be close to the `true' maximum mass conjectured from the GW170817 event.
The two models predict then $R_{1.4}=12.3, 12.7\,$km, respectively.

The new astrophysical constraints on maximum mass and tidal deformability
exclude several models with too small maximum mass and
the DBHF EOS with a too large deformability.
Tightening the lower limit on $\la_{1.4}$
could potentially exclude several other EOSs.

For all examined EOSs
we also confirmed the correlation between the radius or deformability
of a $1.4\,\ms$ NS
and the pressure of betastable matter at about twice normal density.
Weaker correlations were found with the compressibility
of betastable matter at that density.
On the other hand,
we did not find any clear correlations between NS deformability
and properties of symmetric matter at normal density.

\section*{Acknowledgments}

This work is sponsored by
the National Natural Science Foundation of China under Grant
Nos.~11075037, 11975077 and the China Scholarship Council,
File Nos.~201706410092 and 2018 06100066.
Partial support comes also from ``PHAROS", COST Action CA16214.

\begin{table}[t]%...............................................................
%\squeezetable
%\small
%\begin{center}
\caption{
Parameters of the fit for the energy per nucleon $E$,
Eq.~(\ref{e:fitf}),
for symmetric nuclear matter (SNM) and pure neutron matter (PNM)
in two different density domains
and for the different EOSs used.}
\medskip
\def\myc#1{\multicolumn{1}{c}{$#1$}}
\renewcommand{\arraystretch}{1.2}
\renewcommand\tabcolsep{4.6pt}
%\begin{ruledtabular}
\begin{tabular}{lr|rrrc|rrrr}
\hline\hline
\multicolumn{2}{c|}{\multirow{2}{*}{EOS}} &
\multicolumn{4}{c|}{$\rho=(0.08\!-\!1)\fm3$} &
\multicolumn{4}{c}{$\rho=(0.14\!-\!0.21)\fm3$} \\
     &     & \myc{a} & \myc{b} & \myc{c} & \multicolumn{1}{c|}{$d$}
           & \myc{a} & \myc{b} & \myc{c} & \myc{d} \\
\hline
\multirow{2}{*}{BOB} & SNM & -65 & 498 & 2.67 & -9 &-189 & 446 & 1.83 & -0.83 \\
                     & PNM &  57 & 856 & 2.91 &  4 &  15 & 584 & 2.37 &  7.11 \\
\hline
\multirow{2}{*}{V18} & SNM & -60 & 369 & 2.66 & -8 & -82 & 487 & 2.58 & -4.96 \\
                     & PNM &  37 & 667 & 2.78 &  6 &  38 & 578 & 2.67 &  5.88 \\
\hline
\multirow{2}{*}{N93} & SNM & -42 & 298 & 2.61 &-12 & -62 & 803 & 3.20 & -8.18 \\
                     & PNM &  67 & 743 & 2.71 &  4 &  42 & 471 & 2.48 &  5.47 \\
\hline
\multirow{2}{*}{UIX} & SNM &-174 & 323 & 1.61 & -4 & -46 & 926 & 3.38 & -9.29 \\
                     & PNM &  24 & 326 & 2.09 &  6 &  31 & 294 & 2.10 &  6.25 \\
\hline\hline
\end{tabular}
%\end{ruledtabular}
%\footnotetext[1]{}
\label{t:fit}
%\end{center}
\end{table}%....................................................................

%===============================================================================
\section*{Appendix: Parametrizations of the BHF EOSs}
%\appendix{Appendix: Parametrizations of BHF EOS}
\label{s:app}

For convenience we provide here simple parametrizations of our numerical results
for the different EOSs,
namely analytical fits of the energy per nucleon $E$ for SNM and PNM.
We find that in both cases the following functional forms constitute excellent
representations of the numerical values
\beq
 E(\rho) = a \rho + b \rho^c + d \:,
\label{e:fitf}
\eeq
where $E$ and $\rho$ are given in MeV and $\fm3$, respectively.
The parameters of the fits are listed in Table~\ref{t:fit}
for the different EOSs we are using.
We provide two sets of parametrizations, i.e.,
a first set to be used for NS structure calculations
in the density range (0.08--1)$\fm3$,
and a second set for the range (0.14--0.21)$\fm3$,
more appropriate for a precise determination of the saturation properties.
The rms deviations of fits and data are better than 1 MeV / 0.02 MeV
for the two cases and for all EOSs.

For asymmetric nuclear matter,
it turns out that the dependence on proton fraction
$x_p=(1-\delta)/2$
can be very well approximated by a parabolic law as assumed in Eq.~(\ref{e:ea})
\cite{bombaci,uix3},
\beq
 {E}(\rho,\delta) \approx
 \esnm(\rho) + \delta^2 \big[ \epnm(\rho) - \esnm(\rho) \big] \:.
\eeq
Therefore, for the treatment of the asymmetric and beta-stable case,
it is only necessary to provide parametrizations for SNM and PNM.

%===============================================================================
%\section*{Authors contribution}
% The section below may be edited at your convenience to acknowledge
% each author's contribution to the manuscript.
% You may remove it if you are a single author.

%All the authors were involved in the preparation of the manuscript.
%All the authors have read and approved the final manuscript.

% Override the revtex href command in order that the JHEP bib style
% will work properly:
%\renewcommand{\href}[2]{#2}

%===============================================================================
\newcommand{\apjl}{Astrophys. J. Lett.\ }
\newcommand{\apj}{Astrophys. J.\ }
\newcommand{\physrep}{Phys. Rep.\ }
\newcommand{\mnras}{Mon. Not. R. Astron. Soc.\ }
\newcommand{\aap}{Astron. Astrophys.\ }
\newcommand{\prc}{Phys. Rev. C\ }
\newcommand{\prd}{Phys. Rev. D\ }
\newcommand{\prl}{Phys. Rev. Lett.\ }
\newcommand{\nphysa}{Nucl. Phys. A\ }
\newcommand{\plb}{Phys. Lett. B\ }
\newcommand{\epja}{EPJA\ }

\bibliographystyle{epj}
\bibliography{coreos}

\end{document}